# Structural Disorder and Elementary Magnetic Properties of Triangular Lattice ErMgGaO$_4$ Single Crystals

F. Alex Cevallos, Karoline Stolze, and Robert J. Cava

Department of Chemistry, Princeton University, Princeton NJ 08544 USA

*Corresponding author: fac2@princeton.edu (F. Alex Cevallos)

**Abstract**

The single crystal growth, structure, and basic magnetic properties of ErMgGaO$_4$ are reported. The structure consists of triangular layers of magnetic ErO$_6$ octahedra separated by a double layer of randomly occupied non-magnetic (Ga,Mg)O$_5$ bipyramids. The Er atoms are positionally disordered. Magnetic measurements parallel and perpendicular to the *c* axis of a single crystal reveal dominantly antiferromagnetic interactions, with a small degree of magnetic anisotropy. A weighted average of the directional data suggests an antiferromagnetic Curie Weiss temperature of approximately -30 K. Below 10 K the temperature dependences of the inverse susceptibilities in the in-plane and perpendicular-to plane directions are parallel, indicative of an isotropic magnetic moment at low temperatures. No sign of magnetic ordering is observed above 1.8 K, suggesting that ErMgGaO$_4$ is a geometrically frustrated magnet.

1. Introduction

Geometrically frustrated magnets are systems where the structure of the material hinders the appearance of long-range magnetic ordering on cooling to low temperatures[1-3]. The simplest example is the case of antiferromagnetic interactions on an equilateral triangular lattice: as each corner is equidistant from the other two on such a lattice, there is no simple way to arrange magnetic spins such that they are purely antiparallel to their nearest neighbors. Thus the spins may fail to order down to temperatures below the Weiss temperature, and when ordering finally does occur it is frequently into a state that is a compromise between parallel and antiparallel spins[1-6].

The $YbFe_2O_4$ structure type is layered, with a single-layer equilateral triangular lattice of large-metal-oxygen octahedra sharing edges separated by a bilayer of M-O triangular bipyramids[7]. Compounds in this family exhibit a wide variety of unusual electronic and magnetic properties, including ferroelectricity[8-11], charge density waves[9,12], and spin glass behavior[13-16]. As the structure allows for many different kinds atoms to occupy each site, many compounds have been synthesized[17,18]. Recently, the $YbFe_2O_4$-type material $YbMgGaO_4$ has drawn attention as a potential quantum spin liquid candidate[19-24], inviting investigation of the properties of the most closely-related compounds in the family. Here we report the single-crystal growth and basic magnetic properties of $ErMgGaO_4$, a compound whose crystal structure has been reported[17,25] though we report in the following that one significant aspect of the structure was overlooked - the Er site disorder.

2. Materials and Methods

Small crystals of $ErMgGaO_4$ were synthesized via the floating zone method from a stoichiometric mixture of $Er_2O_3$ (99.99%, Alfa Aesar), $Ga_2O_3$ (99.999%, Alfa Aesar), and MgO (99.95%, Alfa Aesar). The powders were ground together in ethanol using an agate mortar and pestle and mixed with stearic acid in a 12:1 molar ratio. When dry, the powder was loaded into a rubber tube and hydrostatically compressed to 40 MPa. The resulting rod (8 cm long, 6 mm diameter) was heated at 1°C per minute to a temperature of 1200°C, heated for four hours, then cooled to room temperature. The rod was then transferred to a four-mirror optical floating zone (FZ) furnace (Crystal Systems, Inc. Model No. FZ-T-10000-HVP-II-P) with 4 x 1500 W lamps. The rod was sintered in the FZ furnace at 50.0% output power with a rotation rate of 10 RPM and a rate of travel of 5 mm/hr. The sintered rod was then cut into lower and upper feed rods, and a crystal growth was performed at 54.1% output power, with a rotation rate of 12 RPM and a growth rate of 3 mm/hr. The resulting crystal growth displayed large, clear facets and had a red color, but was cloudy and translucent when viewed under a microscope, suggesting a high degree of either twinning or impurity phases. Small portions of the crystal facets were carefully removed from the rod, and determined to be clear platelike single crystals of $ErMgGaO_4$. No pure crystals were obtained larger than 1 mm in largest dimension.

The crystal structure of $ErMgGaO_4$ was determined by single-crystal X-ray diffraction (SXRD), especially to test for possible positional disorder of Er and to evaluate the Ga/Mg mixed occupancy position. The SXRD data was collected at 296 K with a Bruker D8 VENTURE diffractometer equipped with a PHOTON CMOS detector using graphite-monochromatized Mo-$K\alpha$ radiation ($\lambda$ = 0.71073 Å). The raw data was corrected for background, polarization, and the Lorentz factor using APEX3 software[26], and multi-scan absorption correction using the SADABS-2016/2 program package was applied[27]. The structure was solved with the charge flipping method[28] and subsequent difference Fourier analyses with Jana2006[29-30]. Structure refinement against $F_o^2$ was performed with Shelxl-2017/1[31-32]. Additional room temperature measurements on single crystal plates lying on a glass slide were taken using a Bruker D8 Advance Eco diffractometer with Cu-$K\alpha$ radiation ($\lambda$ = 1.5418 Å) and a LynxEye-XE detector.

Magnetic measurements were taken using a Quantum Design Physical Property Measurement System (PPMS) Dynacool with a vibrating sample mount. The temperature-dependent DC magnetization (M) was measured for a 1.1 mg (~1 mm x 0.2 mm) single crystal of $ErMgGaO_4$ from 1.8 K to 300 K. The magnetic susceptibility was defined as M/H, where H is the applied field in Oe.

Magnetization vs. applied field measurements in fields of up to 90 kOe were taken at 2 K and 15 K. Both types of measurements were taken both parallel and perpendicular to the *c*-axis of the crystal. For measurements parallel to *c*, the crystal was placed in a plastic sample holder. For measurements perpendicular to *c*, the sample was mounted onto a silica sample holder with GE varnish.

## 3. Results and Discussion

*3.1 Crystal Structure*

The crystal structure of ErMgGaO$_4$ was determined from the refinement of single crystal XRD data and compared to previously published structure models[17,25]; it can be seen in Figure 1.a. ErMgGaO$_4$ crystallizes in the space group $R\bar{3}m$ (no. 166), with ambient temperature lattice parameters were determined to be $a$ = 3.4317(1) Å and $c$ = 25.104(1) Å. Although previous reports suggested that the Er atom sits directly on the high-symmetry [0 0 0] site, the initial structure model with anisotropic refined thermal parameters for Er on this site (Wyckoff site 3*a*, [0 0 0]) revealed an anomalously broad electron distribution along the *z*-axis. The examination of the Fourier ($F_{obs}$) map (Figure 1.b) revealed significantly elongated electron density maxima along *z* for the Er position, clearly reflecting the presence of positional disorder of the rare earth metal. An appropriate displaced site (Wyckoff site 6*c*, with a free parameter allowing the displacement of the Er along *c*) was therefore introduced to describe the statistical positional disorder of the rare earth atom. The refined positional displacement of the Er from [0,0,0] was highly significant, at about 60 standard deviations, with coordinates refined to [0, 0, 0.00418(7)]. 50% occupancy was used for each of these Er sites because the resulting Er positions are separated by 2 x 0.00418 x 25.1 Å = 0.21 Å and are therefore too close to be simultaneously occupied. This position for the rare earth atom, displaced along *c*, is consistent with the published structures of several related compounds[13]. The disorder for the rare earth atom positions in the related materials YbMgGaO$_4$[22] and TmMgGaO$_4$[33] has previously been attributed to the disorder in the local environment caused by the random mixing of Mg and Ga atoms in the neighbouring layers. As is generally the case for mixed atom occupancy positions, the displacement parameters and the coordinates for the Ga and Mg atoms were constrained to be equal in our structural refinement of ErMgGaO$_4$. The final atomic parameters are summarized in Table 1, and further crystallographic data can be seen in the supplementary information (Supplementary Tables 1 and 2).

*3.2 Magnetism*

A platelike single crystal of ErMgGaO$_4$ was isolated from the crystal growth, placed on a glass slide, and analyzed on the powder X-ray diffractometer in the flat plate geometry. This XRD scan showed only peaks corresponding to the (00$l$) planes in the ErMgGaO$_4$ crystal structure, indicating that the large face of the crystal was orthogonal to the $c$-axis. This X-ray pattern, as well as a photo of the thus characterized crystal used in the diffraction and magnetization measurements can be seen in Figure 1.c.

The temperature-dependent magnetic susceptibility of single crystal ErMgGaO$_4$, measured parallel and perpendicular to the $c$-axis in a field of 10 kOe, is shown in Figure 2. The susceptibility was fit to the Curie-Weiss Law, $\chi - \chi_0 = C / (T - \theta_W)$, where $\chi$ is the susceptibility, $C$ is the Curie Constant, $\theta_W$ is the Weiss temperature, and $\chi_0$ is a temperature-independent contribution. This contribution was determined to be -0.004 emu mol$^{-1}$ parallel to $c$, and -0.0055 emu mol$^{-1}$ perpendicular to $c$, values that include the contributions from the sample holder and are therefore not considered precise. The resulting values for the Curie-Weiss fits at high temperature (150-300 K) and low temperature (2-10 K) are summarized in Table 2. For measurements parallel to $c$ at high temperature, $\theta_W$ was determined to be -51.9 K, and C was determined to be 14.21. The effective magnetic moment ($\mu_{eff}$) was determined by the equation $\mu_{eff} = \sqrt{8C}$, and was found to be 10.7 $\mu_B$. Perpendicular to $c$, high-temperature measurements produced values for $\theta_W$ of -18.4 K, $C$ of 8.68, and an effective magnetic moment of 8.3 $\mu_B$, suggesting a degree of magnetic anisotropy that is likely due to the material's layered structure. The negative Weiss temperatures are indicative of dominantly antiferromagnetic exchange interactions, but the susceptibility curves themselves show no indication of magnetic ordering transitions down to 1.8 K. This suggests that ErMgGaO$_4$ is a material that exhibits a significant degree of geometric magnetic frustration, much like the related materials TmMgGaO$_4$[33] and YbMgGaO$_4$[19].

At lower temperatures, the susceptibilities show a notable deviation from ideal Curie-Weiss behavior. The inverse susceptibility curves of the two orientations cross near 20 K, and below 10 K they are almost parallel. Curie-Weiss fits in this low temperature region resulted in values of $C = 5.75$, $\theta_W = -7.5$ K, and $\mu_{eff} = 6.8$ $\mu_B$ parallel to $c$, and values of $C = 6.12$, $\theta_W = -5.3$ K, and $\mu_{eff} = 7.0$ $\mu_B$ perpendicular to $c$.

Field-dependent magnetization was measured along both orientations at 2 K, the results of which can be seen in the inset of Figure 2. The measurement parallel to $c$ showed a linear magnetic response up to approximately $\mu_0 H = 2$ Tesla, although $M$ vs. $H$ curves significantly at higher fields, not showing a tendency towards saturation below $\mu_0 H = 9$ Tesla. Perpendicular to $c$, the magnetization curve shows again shows linearity up to fields of approximately $\mu_0 H = 2$ Tesla, but then shows a greater tendency towards saturation, although saturation is not reached by $\mu_0 H = 9$ Tesla. The magnetization observed near $\mu_0 H = 9$ Tesla is significantly less than what is expected if all the Er moments are oriented along the magnetic field direction. From the magnetic data it can be concluded that ErMgGaO$_4$ exhibits a degree of magnetic anisotropy, although the effect is much weaker than in the closely related system TmMgGaO$_4$[33] and is closer in magnitude to what is observed in YbMgGaO$_4$ and YbZnGaO$_4$[34].

The individual single crystals of ErMgGaO$_4$ used for the diffraction and magnetization measurements were identified via the use of the X-ray Diffractometer in flat plate geometry to look at the out-of-plane X-ray scattering, as described above. The mixing of such crystals with impurity phases such as the Er$_3$Ga$_5$O$_{12}$ garnet in the floating zone growth rod made it impractical to isolate a bulk powder sample for magnetic characterization, and thus we estimate the average magnetic behavior by taking the averages of the directional measurements on the single crystal (2/3 of the susceptibility perpendicular to $c$, and 1/3 of the susceptibility parallel to $c$). The resulting curve is plotted in Figure 3.

A high-temperature (150-300 K) Curie-Weiss fit of this average curve yielded values of $C = 10.49$, $\theta_W = -30.4$, and $\mu_{eff} = 9.2$ $\mu_B$. This value for the effective magnetic moment is close to the ideal of 9.59 $\mu_B$ per Er expected for a free ion. At low temperatures, the average curve is nearly parallel to the two directional measurements, as expected from the direction-dependent data. A Curie-Weiss fit to the region between 2 K and 10 K yielded values of $C = 6.01$, $\theta_W = -5.9$, and $\mu_{eff} = 6.9$ $\mu_B$. A temperature-dependent magnetization measurement in the low temperature region in an applied field of 1 kOe is displayed in the rightmost inset in Figure 3, and shows that in this lower field, there is also no evidence of magnetic ordering in the magnetic susceptibility down to temperatures of 1.8 K.

## 4. Conclusions

Small single crystals of the equilateral triangular plane lattice material ErMgGaO$_4$ have been synthesized via the optical floating zone method, and the structure has been determined using single-crystal X-ray diffraction. The lattice parameters and atomic positions are generally found to be in good agreement with previously published values for this and other isostructural compounds. Notably, however, the magnetic Er ions are in disordered positions, a consideration that is important when modeling the magnetic behavior. Magnetic measurements suggest that the magnetism in the material is dominated by antiferromagnetic interactions, although no magnetic ordering is observed above 1.8 K, suggesting the presence of geometric magnetic frustration. Some anisotropy is observed in the magnetic measurements, in line with related compounds in this structure type and consistent with the layered structure of the material.

Continued work on the synthesis of this material is warranted. A pure powder sample may be of interest for initial characterization of the magnetic system by neutron diffraction. Additionally, this material and its family of related compounds is worthy of further study due to its combination of an isolated, equilateral triangular plane lattice of spins plus subtle structural distortions of that lattice caused by local disorder in non-magnetic ions in the adjacent planes. Finally, the growth of larger single crystals for detailed neutron scattering measurements may provide valuable insight into these compounds, and warrants further effort.

**Acknowledgments**

This research was supported by the US Department of Energy, Division of Basic Energy Sciences, Grant No. DE-FG02-08ER46544, and was performed under the auspices of the Institute for Quantum Matter.

**Table 1.** Wyckoff positions, coordinates, occupancies, and equivalent isotropic displacement parameters respectively for an ErMgGaO$_4$ single-crystal measured at 296(1) K. $a$ = 3.4317(1) Å, $c$ = 25.104(1) Å. The coordinates and thermal parameters of Ga and Mg were constrained to be equal; $U_{eq}$ is one third of the trace of the orthogonalized $U_{ij}$ tensor.

| Atom | Wyck. Site | $x$ | $y$ | $z$ | Occupancy | $U_{eq}/U_{iso}$ |
|---|---|---|---|---|---|---|
| Er | 6$c$ | 0 | 0 | 0.00418(7) | 0.5 | 0.00721(18) |
| Ga | 6$c$ | 0 | 0 | 0.21406(2) | 0.5 | 0.00484(5) |
| Mg | 6$c$ | 0 | 0 | 0.21406(2) | 0.5 | 0.00484(5) |
| O1 | 6$c$ | 0 | 0 | 0.29068(6) | 1 | 0.00751(17) |
| O2 | 6$c$ | 0 | 0 | 0.12908(7) | 1 | 0.0131(2) |

**Table 2.** Magnetic properties of ErMgGaO$_4$, as determined from magnetization vs. temperature measurements parallel and perpendicular to the $c$-axis in an applied field of 10 kOe.

| Direction | Region of M-T curve | Curie Constant ($C$) | Weiss Temperature ($\theta_W$) | Effective Magnetic Moment ($\mu_{eff}$) |
|---|---|---|---|---|
| Parallel to $c$ | High-T | 14.21 | -51.9 K | 10.7 $\mu_B$ |
|  | Low-T | 5.75 | -7.5 K | 6.8 $\mu_B$ |
| Perpendicular to $c$ | High-T | 8.68 | -18.4 K | 8.3 $\mu_B$ |
|  | Low-T | 6.12 | -5.3 K | 7.0 $\mu_B$ |
| Weighted average | High-T | 10.49 | -30.4 K | 9.2 $\mu_B$ |
|  | Low-T | 6.01 | -5.9 K | 6.9 $\mu_B$ |

**Figure captions**

**Figure 1. (a)** The crystal structure of ErMgGaO$_4$ showing the coordination polyhedra. ErO$_6$ octahedra are olive, Mg/GaO$_5$ triangular bipyramids are green. On the right is the triangular arrangement of Er atoms as viewed down the $c$-axis. In the bottom right is a closer view of the distorted ErO$_6$ octahedra and the slightly off-center nature of the Er positions. **(b)** $F_{obs}$ Fourier map for ErMgGaO$_4$ based on room-temperature data; in space group $R\bar{3}m$, map is sum of electron density between $-0.10 < y < 0.10$, contour lines correspond to 12 e/Å$^3$. **(c)** X-ray diffraction pattern of a single crystal of ErMgGaO$_4$, showing only peaks corresponding to (00$l$) planes. Inset: A picture of the actual crystal used in measurements.

**Figure 2.** Temperature-dependent magnetic susceptibility of ErMgGaO$_4$ measured parallel (blue) and perpendicular (red) to the $c$-axis in an applied field of 10 kOe. Inset: The field-dependent magnetization of ErMgGaO$_4$ measured parallel and perpendicular to the $c$-axis at a temperature of 2 K.

**Figure 3**. Temperature-dependent inverse susceptibility curves of ErMgGaO$_4$ measured parallel (blue) and perpendicular (red) to the $c$-axis in an applied field of 10 kOe. In pink is the weighted average of the two curves (2/3 perpendicular, 1/3 parallel). Left inset: The low-temperature region of the temperature-dependent inverse susceptibility curves. Right inset: The low-temperature region of the temperature-dependent susceptibility curves measured in an applied field of 1 kOe.

**Figures**

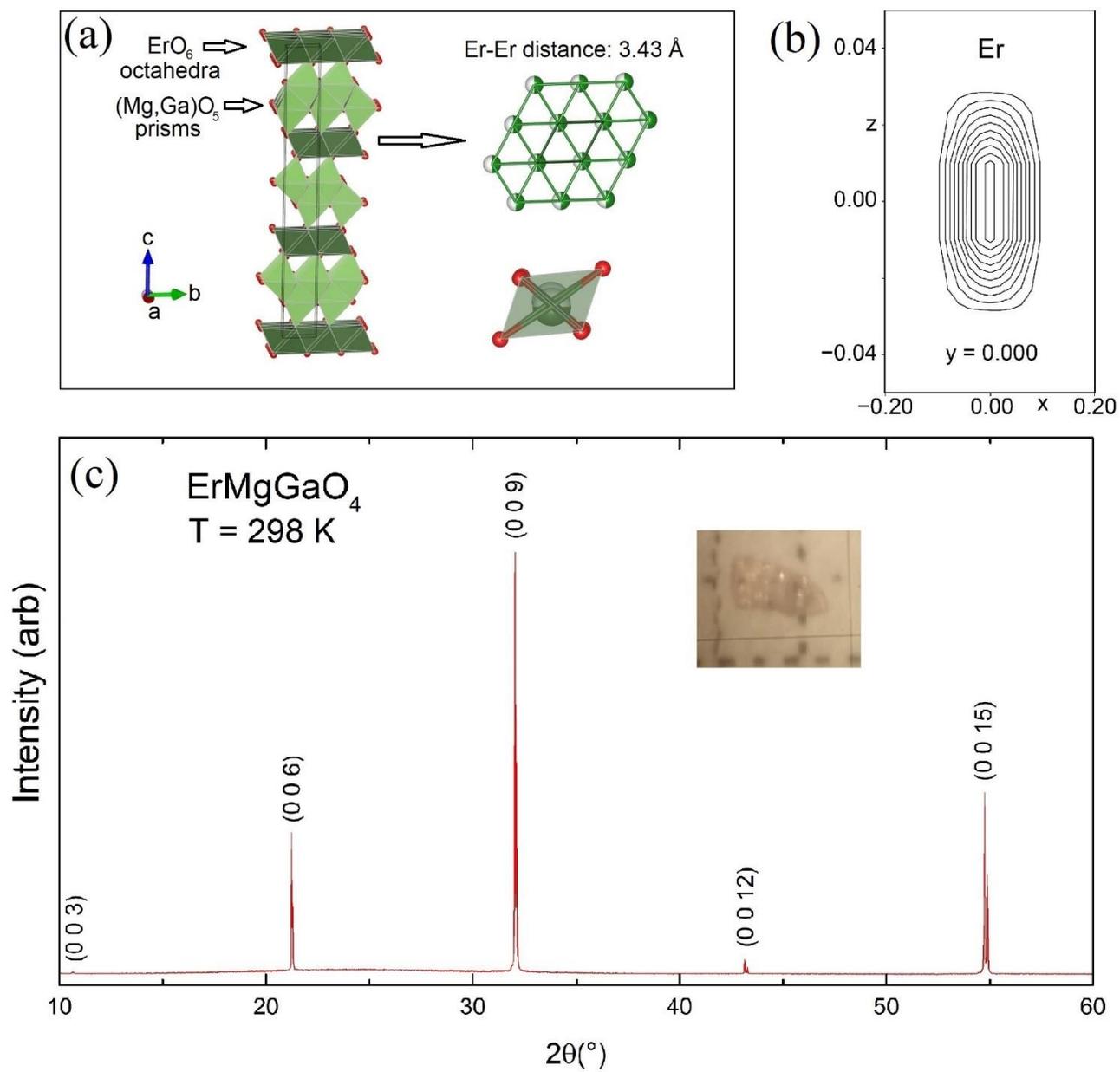

Figure 1

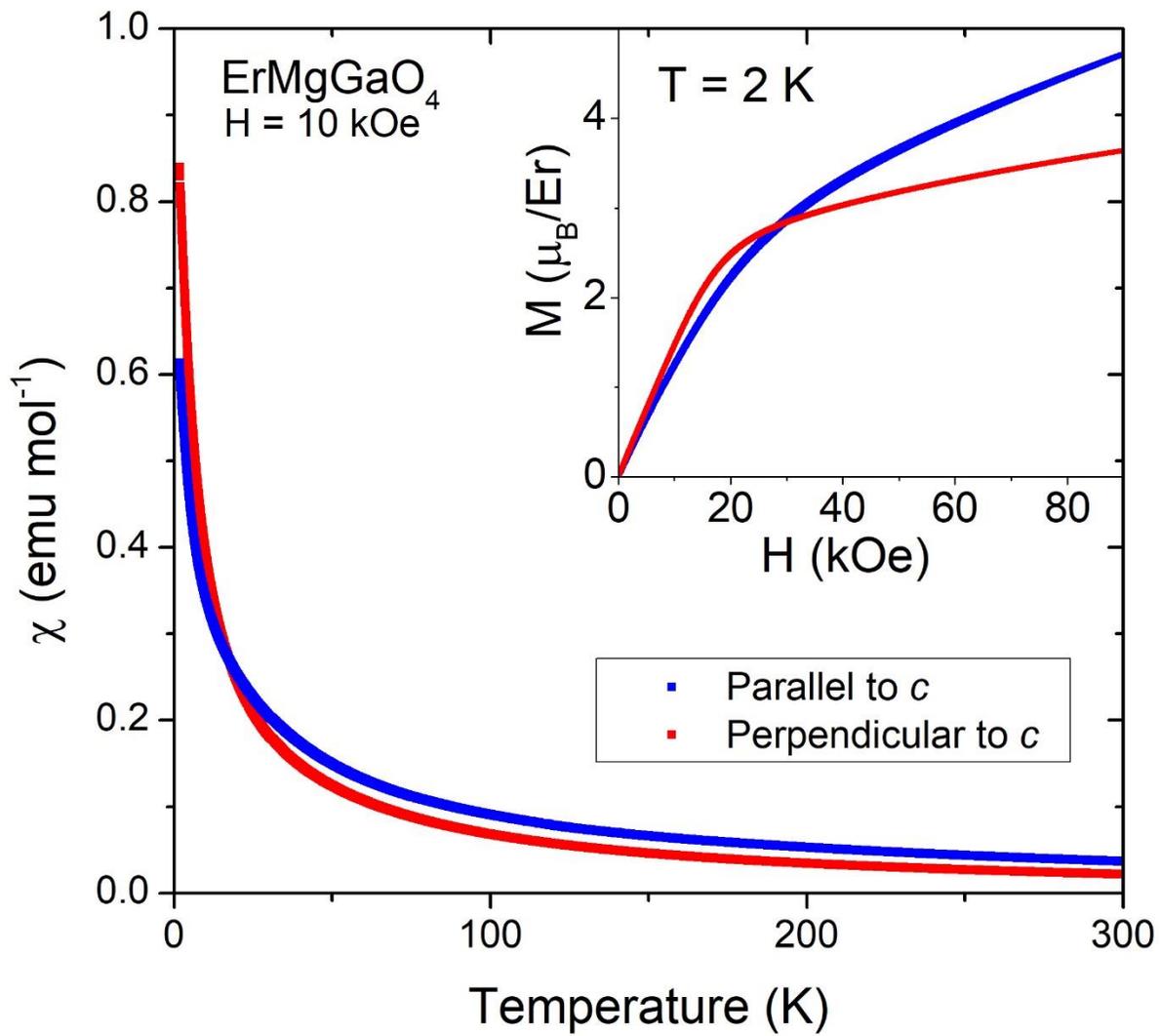

Figure 2

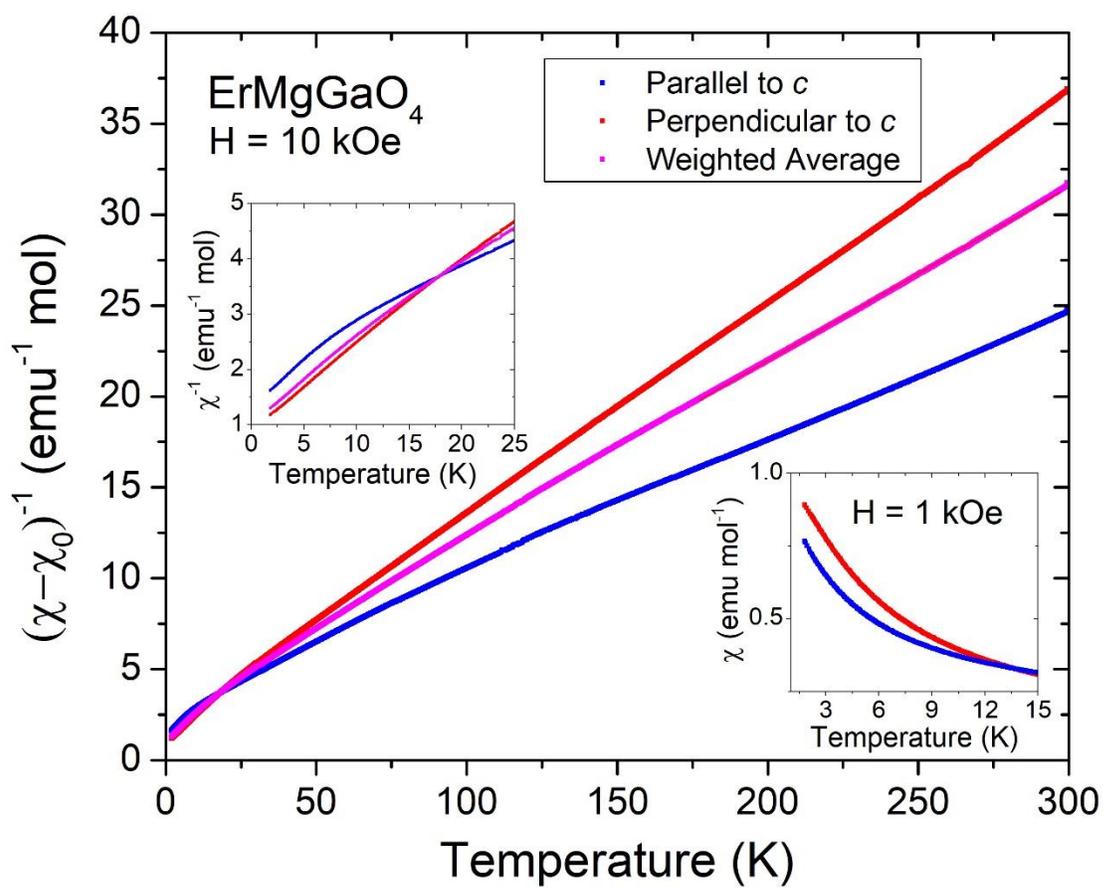

Figure 3

**Supplementary Information**

**Supplementary Table 1.** Crystallographic data and details of the structure determination of ErMgGaO$_4$ derived from single-crystal experiments measured at 296(1) K.

| | |
|---|---|
| Sum Formula | ErMgGaO$_4$ |
| Formula weight / (g · mol$^{-1}$) | 325.29 |
| Crystal System | trigonal |
| Space group | *R-3m* (no. 166) |
| Formula units per cell, Z | 3 |
| Lattice parameter *a* / Å | 3.4317(1) |
| *c* / Å | 25.104(1) |
| Cell volume / (Å$^3$) | 256.03(2) |
| Calculated density / (g · cm$^{-3}$) | 6.329 |
| Radiation | $\lambda$(Mo-*K$\alpha$*) = 0.71073 Å |
| | $2\theta \leq 95.66°$ |
| Data range | $-7 \leq h \leq 7$ |
| | $-7 \leq k \leq 7$ |
| | $-50 \leq l \leq 51$ |
| Absorption coefficient / mm$^{-1}$ | 32.33 |
| Measured reflections | 6738 |
| Independent reflections | 362 |
| Reflections with $I > 2\sigma(I)$ | 345 |
| *R*(int) | 0.036 |
| *R*($\sigma$) | 0.013 |
| No. of parameters | 12 |
| $R_1$(obs) | 0.011 |
| $R_1$(all $F_o$) | 0.012 |
| $wR_2$(all $F_o$) | 0.023 |
| Residual electron density / (e · Å$^{-3}$) | 0.95 to –0.94 |

**Supplementary Table 2.** Anisotropic thermal displacement parameters for ErMgGaO$_4$ single-crystal measured at 296(1) K. The coefficients $U_{ij}$ (/Å$^2$) of the tensor of the anisotropic temperature factor of atoms are defined by $\exp\{-2\pi^2[U_{11}h^2a^{*2} + \cdots + 2U_{23}klb^*c^*]\}$; $U_{ij}$[Ga] and $U_{ij}$[Mg] were constrained to be equal.

| Atom | $U_{11}$ | $U_{22}$ | $U_{33}$ | $U_{12}$ |
|---|---|---|---|---|
| Er | 0.00290(4) | 0.00290(4) | 0.0158(5) | 0.00145(2) |
| Ga/Mg | 0.00448(6) | 0.00448(6) | 0.00557(12) | 0.00224(3) |